\documentclass[12pt]{article}
\usepackage{amssymb,amsfonts}
\usepackage{epsf,epsfig}
\begin{document}
\baselineskip 18pt

\title{A model with generalized Y-junction}
\author{P.~N.~Bibikov and L.~V.~Prokhorov}

\maketitle

\vskip5mm

\begin{abstract}
The Klein-Fock-Gordon equation is studied on the generalized
Y-junction of $N$ strings with a massive center. The corresponding formulas for wave
scattering and normal modes are obtained.
\\
\\
{\bf Key words} differential equations on networks, transmission
rules, Klein-Fock-Gordon equation.
\end{abstract}

\section {Introduction.}

Dynamical systems on manifolds are standard objects studied both
in classical and quantum mechanics. Here we continue to study
mechanics not on manifolds. The simplest example of such a system
gives a 3-ray star studied in [1] (see also [2] and [3]). This
space is not a manifold. The problem became of special importance
a couple decades ago after the discovery of D-branes. Problems of
this type arise in different
branches of Physics.\\
{\it String theory.} Effectively strings are ordered sets of
interacting harmonic oscillators and end of a string can belong to
another string
(D1-brane). Problem: describe the evolution of excitations of the system. \\
{\it Nanoelectronics and theory of polymers.}
Nanoelectronics and theory of polymers are important for modern technologies.\\

The importance of this problem for strings is self evident.
Gradually it becomes clear that at the Planck scale matter
manifests itself in form of strings, and a 3D-network of
superstrings can model the physical space [4], clearly leading to
unification of all interactions, including gravitation.

There are two types of problems (both in classical and quantum
mechanics): (i) determination of normal coordinates, (ii)
description of scattering  (i.e. description of evolution of waves
given on the rays). It turns out that here normal coordinates are
given by linear transformations, i.e. these systems in fact give
examples of free field theories. The relativistic ${\rm Schr\ddot
odinger}$ equation on a tail (i.e. the Klein-Fock-Gordon equation)
coincides with the equation of motion of a free field. Thus the
problems of scattering on the junction are in fact identical both
in classical and quantum cases. The only difference is that in
classical physics one usually considers scattering of a wave
packet, while in quantum mechanics it is the scattering of a
particle with certain momentum $k$. It is significant that though
the fields are formally free there exists nontrivial scattering: a
wave on a ray may either be reflected or pass to an other tail
[1].

A bosonic string is in fact an ordered set of harmonic oscillators
with identical masses. In nanoelectronics there are 3-ray stars
made of 3 quantum wires and a quantum well (or gate). The latter is modeled
by a central oscillator. That is why here we study the case of a
nontrivial junction when the mass of the central oscillator is
different from the mass of the others. The problem of the 2-ray
star (oscillators on an axis), one oscillator with an arbitrary
mass, was studied in [5]. This problem is close to the problem
of one-dimensional scattering on a $\delta$-potential.

We solve here the problem of scattering on such a junction in both
classical and quantum cases (Sec. 2) and find normal coordinates
in the general case ($N$-tail star) (Sec. 3).

\section{Harmonic network approximation}
The Klein-Fock-Gordon equation in 1+1 dimensions has a natural
discretization,
\begin{equation}
\ddot\varphi_n=\frac{1}{\Delta^2}\Big(\varphi_{n+1}+\varphi_{n-1}-2\varphi_n\Big)-m^2\varphi_n.
\end{equation}
Here $\Delta$ is the lattice constant.

A solution of this system of differential equations corresponding
to a monochromatic wave with positive energy has the following
form,
\begin{equation}
\varphi_n(t)={\rm e}^{-i(\omega_k t-kn\Delta)},
\end{equation}
where the dispersion,
\begin{equation}
\omega^2_k=\frac{4}{\Delta^2}\sin^2\frac{k\Delta}{2}+m^2,
\end{equation}
in the limit $\Delta\rightarrow0$ pass to the well-known
relativistic form.

The simplest way to describe the junction is to introduce for it a
corresponding new variable which is denoted by $u$. If
$\varphi^{(j)}_n$, $j=1,2,3,$ are components of $\varphi$ related
to the corresponding rays we may propose the following Lagrangian,
\begin{equation}
L=\sum_j\Big(L^{(j)}-\frac{1}{2\Delta^2}(u-\varphi^{(j)}_1)^2\Big)+\frac{1}{2}\Big(M\dot
u^2- m^2u^2\Big),
\end{equation}
where
\begin{equation}
L^{(j)}=\frac{1}{2}\sum_{n=1}^{\infty}\Big(\dot\varphi_n^{(j)2}-\frac{1}
{\Delta^2}(\varphi^{(j)}_{n+1}- \varphi^{(j)}_n)^2-
m^2\varphi^{(j)2}_n\Big).
\end{equation}

The Lagrangian (4), (5) produces a nontrivial realization of a
one-dimensional harmonic lattice. It gives the following
equations of motion:
\begin{eqnarray}
M\ddot u&=&\frac{1}{\Delta^2}\sum_j\Big(\varphi^{(j)}_1-u\Big)-m^2u,\\
\ddot\varphi^{(j)}_1&=&\frac{1}{\Delta^2}\Big(\varphi^{(j)}_2+u-2\varphi^{(j)}_1\Big)-
m^2\varphi^{(j)}_1,\\
\ddot\varphi^{(j)}_n&=&\frac{1}{\Delta^2}\Big(\varphi^{(j)}_{n+1}+\varphi^{(j)}_{n-1}-2
\varphi^{(j)}_n\Big)-m^2\varphi^{(j)}_n,\quad n>1.
\end{eqnarray}

Here $\Delta$ is the lattice parameter while $m$ is the
quasiparticle (excitation) mass. The parameter $M$ is
dimensionless.

We now study scattering on the junction which is described by
the following general solution [1],
\begin{eqnarray}
\varphi^{(x)}_n(t)&=&{\rm e}^{-i(\omega_kt+kn)}+R(k){\rm
e}^{-i(\omega_kt-k\Delta n)},\\
\varphi^{(y)}_n(t)=\varphi^{(z)}_n(t)&=&(R(k)+1){\rm e}^{-i(\omega_kt-k\Delta n)},\\
u(t)&=&(R(k)+1){\rm e}^{-i\omega_kt}.
\end{eqnarray}
Our task is to obtain $R(k)$. The Eqs. (6)-(7) lead to
\begin{equation}
2M(1-\cos k\Delta)(R(k)+1)=3(R(k)+1)-{\rm
e}^{-ik\Delta}-(3R(k)+2){\rm e}^{ik\Delta},
\end{equation}
or
\begin{equation}
R(k)=\frac{1}{3}{\rm e}^{i\theta(k)}-\frac{2}{3},
\end{equation}
where
\begin{equation}
{\rm
e}^{i\theta(k)}=-\frac{(2M-3)\sin\frac{k\Delta}{2}-3i\cos\frac{k\Delta}{2}}
{(2M-3)\sin\frac{k\Delta}{2}+3i\cos\frac{k\Delta}{2}}.
\end{equation}

Now we may suppose that the parameter $M$ is a function of
$\Delta$ and take the limit $\Delta\rightarrow0$. Defining a new
parameter
\begin{equation}
k_1=\lim_{\Delta\rightarrow0}\frac{3i}{\Delta\cdot
M(\Delta)}
\end{equation}
we obtain the following equation,
\begin{equation}
{\rm e}^{i\theta(k)}=-\frac{k-k_1}{k+k_1}.
\end{equation}
Eq. (16) is the main result of this section.

\section{Normal modes for the N-star model}
In the discrete model (4) we put for simplicity $\Delta=1$.
This is equivalent to the normalization $|k|\leq\pi$ of the wave
number. Now $j=1,2,...,N.$

The complex coordinates,
\begin{equation}
\varphi^{(j)}(k)=\sum_{n=1}^{\infty}{\rm e}^{ikn}\varphi^{(j)}_n,
\end{equation}
satisfy the following equations:
\begin{equation}
\ddot \varphi^{(j)}(k)=-\omega_k^2\varphi^{(j)}(k)+u{\rm
e}^{ik}-\varphi^{(j)}_1,
\end{equation}
It is convenient to pass to their real and imaginary parts,
\begin{equation}
\varphi_c^{(j)}(k)=\sum_{n=1}^{\infty}\varphi_n^{(j)}\cos
kn,\qquad
\varphi_s^{(j)}(k)=\sum_{n=1}^{\infty}\varphi_n^{(j)}\sin kn.
\end{equation}
They satisfy the equations:
\begin{eqnarray}
\ddot \varphi_c^{(j)}(k)&=&-\omega_k^2\varphi_c^{(j)}(k)+u\cos
k-\varphi_1^{(j)},\\
\ddot \varphi_s^{(j)}(k)&=&-\omega_k^2\varphi_s^{(j)}(k)+u\sin k.
\end{eqnarray}

If we define
\begin{equation}
\xi_0(k)=\sum_{j=1}^{N}\varphi_c^{(j)}(k)+Mu,
\end{equation}
then
\begin{equation}
\ddot\xi_0(k)=-\omega_k^2\xi_0(k)-(N-2)u(1-\cos k).
\end{equation}

The normal modes are:
\begin{equation}
\xi_j(k)=\sin k\xi_0(k)+[(N-2M)(1-\cos
k)+(1-M)m^2]\varphi_s^{(j)}(k).
\end{equation}
They satisfy the following system of equations:
\begin{equation}
\ddot\xi_j(k)=-\omega_k^2\xi_j(k).
\end{equation}

In the case $M=1$ we may divide (24) by $2\sin{k/2}$
and come to a simpler formula,
\begin{equation}
\xi_j(k)=\cos\frac{k}{2}\xi_0(k)+(N-2)\sin\frac{k}{2}\varphi_s^{(j)}(k).
\end{equation}
In this case we may express $u$ and $\varphi_n^{(j)}$ from
$\xi_m(k)$.

\section{Inverse transformation from normal modes for $M=1$}

In order to express $u$ and $\varphi_n^{(j)}$ from $\xi_j(k)$ we
note that according to (24) for $M=1$,
\begin{eqnarray}
\sum_{j=1}^N\xi_j(k)&=&Nu\cos\frac{k}{2}\nonumber\\
&+&\sum_{n=1}^{\infty}\sum_{j=1}^N\varphi_n^{(j)}\Big(N\cos
kn\cos\frac{k}{2}+
(N-2)\sin kn\sin\frac{k}{2}\Big),\\
\xi_j(k)-\xi_{j-1}(k)&=&\sin\frac{k}{2}\sum_{n=1}^{\infty}(\varphi_n^{(j)}-\varphi_n^{(j-1)})\sin
kn.
\end{eqnarray}

Defining new variables,
\begin{equation}
Q_0=Nu,\qquad Q_n=\sum_{j=1}^N\varphi_n^{(j)},\quad n=1,...,
\end{equation}
one may rewrite Eq. (27) in the following form:
\begin{equation}
\sum_{j=1}^N\xi_j(k)=\sum_{n=0}^{\infty}(Q_n+(N-1)Q_{n+1})\cos
k\Big(n+\frac{1}{2}\Big).
\end{equation}
Then according to (28) and (30)
\begin{eqnarray}
\Delta Q_{j,n}
&=&\frac{1}{\pi}\int_{-\pi}^{\pi}\frac{\xi_{j+1}(k)-\xi_j(k)}
{\sin\frac{\displaystyle k}{\displaystyle2}}\sin{kn}dk,\\
\eta_n&=&Q_n+(N-1)Q_{n+1},
\end{eqnarray}
where
\begin{equation}
\Delta Q_{j,n}=\varphi_n^{(j+1)}-\varphi_n^{(j)}.
\end{equation}
and
\begin{equation}
\eta_n=\frac{1}{\pi}\int_{-\pi}^{\pi}\sum_{j=1}^N\xi_j(k)\cos
k\Big(n+\frac{1}{2}\Big)dk.
\end{equation}

The recurrent system (32) may be represented in the following
matrix form:
\begin{equation}
\left(\begin{array}{c}
\eta_1\\
\eta_2\\
...
\end{array}\right)=A
\left(\begin{array}{c}
Q_1\\
Q_2\\
...
\end{array}\right),
\end{equation}
where
\begin{equation}
A=\left(\begin{array}{ccccc}
1&N-1&0&0&...\\
0&1&N-1&0&...\\
0&0&1&N-1&...\\
...&...&...&...&...
\end{array}\right).
\end{equation}
It can be easily proved that
\begin{equation}
A^{-1}=\left(\begin{array}{ccccc}
1&1-N&(1-N)^2&(1-N)^3&...\\
0&1&(1-N)&(1-N)^2&...\\
0&0&1&(1-N)&...\\
...&...&...&...&...
\end{array}\right)
\end{equation}
and from Eqs. (35) and (37) it follows that
\begin{equation}
Q_n=\sum_{m=n}^{\infty}\eta_m(1-N)^m
\end{equation}
or, according to (34),
\begin{equation}
Q_n=\frac{1}{\pi}\int_{-\pi}^{\pi}\sum_{j=1}^N\xi_j(k)\sum_{m=0}^{\infty}[(1-N)^m\cos
k(n+m+\frac{1}{2})]dk.
\end{equation}
Using an equality
\begin{equation}
\sum_{m=0}^{\infty}(1-N)^m\cos k(n+m+\frac{1}{2})={\rm
Re}\frac{{\rm exp}(ik(n+\frac{1}{2}))}{1+(N-1)\exp(ik)}
\end{equation}
one can readily obtain
\begin{equation}
Q_n=\frac{1}{\pi}\int_{-\pi}^{\pi}\sum_{j=1}^N\xi_j(k)\frac{\cos
k(n+\frac{1}{2})+(N-1)\cos
k(n-\frac{1}{2})}{N^2-4(N-1)\sin^2\frac{k}{2}}dk.
\end{equation}

Now according to (29) and (33) one can readily write the
transverse transformation
\begin{eqnarray}
\varphi_n^{(j)}(k)&=&\frac{1}{N}\Big(Q_n-\sum_{m=0}^{N-1}(N-1-m)\Delta
Q_{j+m,n}\Big).
\end{eqnarray}

\section{Conclusions}

In the present paper we have considered the discrete version the
Klein-Fock-Gordon equation in the $Y$-junction with arbitrary mass
of the central oscillator. We obtained the corresponding formulas
for wave propagation. The normal modes were obtained
in the general case of $N$-rays. In the special case when the
junction point mass is equal to unity the explicit formulas for
inverse transformations were also obtained.

The authors are grateful to B. S. Pavlov for his interest to
the paper.

\end{document}